\documentclass[12pt]{elsart}

\usepackage{graphicx}
\usepackage{amsmath}
 \usepackage{color}
\usepackage{changepage}
\usepackage{subfigure}
\usepackage{multirow}

\usepackage{amssymb}
\usepackage [latin1]{inputenc}
\usepackage[compress]{cite}

\begin{document}

\begin{frontmatter}

\title{First encounters on Bethe Lattices and Cayley Trees}
\author[lable1,label2]{Junhao Peng}
\ead{pengjh@gzhu.edu.cn}
\author[label3,label4,label5]{Trifce Sandev}
\ead{trifce.sandev@manu.edu.mk}
\author[label3,label6]{Ljupco Kocarev}
\address[lable1]{School of Math and Information Science, Guangzhou University, Guangzhou 510006, China.}
\address[label2]{Guangdong Provincial Key Laboratory co-sponsored by province and city of Information Security Technology, Guangzhou University, Guangzhou 510006,  China.}
\address[label3]{Research Center for Computer Science and Information Technologies, Macedonian Academy of Sciences and Arts, Bul. Krste Misirkov 2, 1000 Skopje, Macedonia.}
\address[label4]{Institute of Physics \& Astronomy, University of Potsdam, D-14776 Potsdam-Golm, Germany.}
\address[label5]{Institute of Physics, Faculty of Natural Sciences and Mathematics, Ss.~Cyril and Methodius University, Arhimedova 3, 1000 Skopje, Macedonia.}
\address[label6]{Faculty of Computer Science and Engineering, Ss. Cyril and Methodius University, \\ P.O. Box 393, 1000 Skopje, Macedonia}
\begin{abstract}
In this work we consider the first encounter problems between a fixed and/or mobile target A and a moving trap B on Bethe Lattices and Cayley trees. The survival probability (SP) of the target A on the both kinds of structures are analyzed analytically and compared. On Bethe Lattices, the results show that the fixed target will still prolong its survival time, whereas, on Cayley trees, there are some initial positions where the target should move to prolong its survival time. The mean first encounter time (MFET) for mobile target A is evaluated numerically and compared with the mean first passage time (MFPT) for the fixed target A. Different initial settings are addressed and clear boundaries are obtained. These findings are helpful for optimizing the strategy to prolong the survival time of the target or to speed up the search process on Cayley trees, in relation to the target's movement and the initial position configuration of the two walkers. We also present a new method, which uses a small amount of memory, for simulating random walks on Cayley trees.

\begin{keyword}
Random walks \sep survival probability \sep mean first encounter time \sep Cayley trees
\PACS  05.40.Fb,  05.60.Cd
\end{keyword}
\end{abstract}
\end{frontmatter}

\section{Introduction}
\label{intro}
Encounters between target A and a moving trap B on appropriate structures have wide applications in target search~\cite{Rupprecht-Benichou-2016-PRE, Sims-Southall-Humphries-Hays-2009-Nature, Volchenkov2011-cnsns, Viswanathan-Buldyrev-Havlin-Luz-Raposo-Stanley-1999-Nature, Magdziarz-Zorawik-2017-cnsns} and prey-predator models~\cite{Luca2012-PRL}. Related researches often focus on the survival probability of the target and the mean first encounter time (MFET) between the walkers~\cite{Koza-Zbi-1998-PRE, Vot-Escuder-2018-PRE-Encounter, Forrester-1999-JPA-Probability, Yuste-Oshanin-2008-PRE, Oshanin-Vasilyev-2009-PNAS, Szabo-1988-PRL-Diffusion, Schehr-2013-JSP-Reunion, Campari-Cassi-2012-PRE,  George-Patel-2018-XXX}. By comparing the survival probability (SP) or the MFET for a mobile target with SP or the mean first-passage time (MFPT) for immobile target, one can find the effect of the target's move on the search efficiency. In general, the survival probability of a mobile target is less than or equal to the SP of an immobile target in the presence of randomly moving traps~\cite{Moreau2003-PRE-Pascal, Chen-Sun-2012-XXX, Tejedor-2011-JPA-Encounter, Holcman-Kupka-2009-JPA}. In other words, the move of the target would speed up the encounter between the target and the moving traps. This argument is also known as ``Pascal principle''. However, the argument is just proved on $d$-dimension lattice~\cite{Moreau2003-PRE-Pascal, Chen-Sun-2014-JTP} and there are also some different findings. Recent results show, if two walkers start from the same site and the initial 
 position is randomly drawn from the stationary distribution, the move of the target A has no effect on the MFET on finite non-bipartite connected graphs, and the move of the target A fastens the encounter between the two walkers on finite bipartite connected graphs~\cite{Peng_Elena-PRE-2019}. On infinite comb, researchers find different results~\cite{Campari-Cassi-2012-PRE, Chen-Chen-2011-EJP,  Benichou-2015-PRL-Diffusion, Chen-Bei-2008-SPL}, i.e., the ultimate SP for a mobile target is greater than the ultimate SP for an immobile target in the presence of a randomly moving trap. On finite comb, the move of the target A can speed up the encounter process on some initial position settings, and the move of the target A can also slow the encounter process on some other initial position settings~\cite{Elena-2014-PRE, Agliari-2016-PRE-Two, Peng_Elena-PRE-2019}. Therefore, one can optimize the search strategy in respect to the target movement and the initial position configuration.

Bethe Lattices are infinite trees with same degree of all vertexes, and Cayley trees are finite trees with same degree of all non-leaf vertexes. Cayley trees can also be looked as Bethe Lattices confined in finite environments. Random walks and encounters on Bethe Lattices and Cayley trees have attracted lots of attention, and there are results for immobile target A~\cite{Ostilli2012-Physica-A, Cassi_1989-EPL, Wu-Lin-Zhang-2012, Lin-Zhang-JCP-2013}. Random walks on Bethe Lattices are non-recurrent, whereas random walks on Cayley trees are recurrent. What is the effect of target's motion on the encounter for two particles walking on Bethe Lattices and Cayley trees? What is the difference between encounters on Bethe Lattices and those on Cayley trees? Are Bethe Lattices thermodynamic limit of Cayley trees? They are interesting problems which are unsolved.

In this paper, we focus on encounters between two walkers (i.e., target A and trap B) performing simple random walk on Bethe Lattices and Cayley trees. On Bethe Lattices, there is a probability that the two walkers will never encounter and there is a nonzero probability that the target will survive forever. We will consider the SP of a randomly moving target A in the presence of randomly moving trap B and compare it with the SP for immobile target A. On the finite Cayley tree structure, the two walkers will encounter in any case whether the target A is mobile or not. In addition to the analysis of SP for the target A, we also evaluate the MFET and compare it with MFPT. Here we address the MFET numerically and different initial position settings are considered. Note that shortage of memory is a common difficult for simulating random walk on graph with large size $N$. Here we introduce a new method which just need $O(log(N))$ memory units. Therefore, the difficulties in memory shortage for numerical simulation are solved.

This paper is organized as follows. In Sec.~\ref{sec:Network} we describe the topology of Bethe lattices and Cayley trees. Next, in Sec.~\ref{sec: SP}, we analyze analytically the SP for the target A on Bethe Lattices and Cayley trees. In Sec.~\ref{sec: FET}, the MFET for mobile target A is evaluated numerically and compared it with the MFPT for immobile target A. Finally, conclusions and discussions are provided in Sec.~\ref{sec:Conclusion}.

\section{The network structure}
\label{sec:Network}

\begin{figure}
\begin{center}
\includegraphics[scale=0.85]{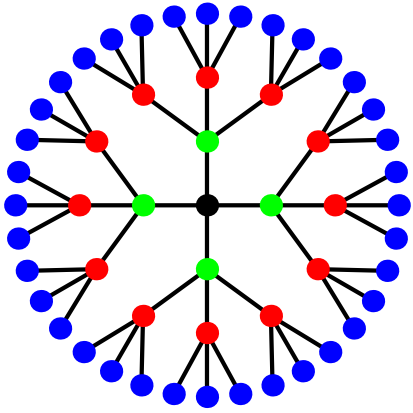}
\caption{Structure for a particular Cayley tree $C_{4,3}$. Vertices colored with green belong to the first shell, with red to the second shell, and with blue to the third shell of $C_{4,3}$. 
}
\label{fig:1}
\end{center}
\end{figure}

Both, the Bethe lattice and Cayley tree, are simple connected undirected graphs $G = (V, E)$ ($V$ represents the set of vertices and $E$ represents the set of edges) with no cycles, and they both are trees. The Bethe lattice, denoted by $BL_m$, is an infinite tree with same degree $m$ $(m\geq 3)$ of all vertices, where $m$ is called a coordination number. Actually, the Bethe Lattice is an unrooted tree, since any vertex will serve equally well as a root. The Cayley tree is rooted, all other nodes are arranged in shells around its root vertex~\cite{Ostilli2012-Physica-A, Cassi_1989-EPL}, and each non-terminal vertex is connected to $m$ $(m\geq 3)$ neighbors. A Cayley tree with coordination number $m$ and $g$ $(g\geq 0)$ shells, $C_{m,g}$, is defined in the following way. We link the root vertex $O$ with $m$ new vertices by means of $m$ edges. The first set of $m$ nodes constitutes the shell $k=1$ of $C_{m,g}$. Then, to build the shell $k$ ($2 \leq k \leq g$), each vertex of the shell $k-1$ is linked to $m-1$ new vertices. The set of these new vertices constitutes the shell $k$ of $C_{m,g}$. FIG.~\ref{fig:1} shows the construction of the Cayley tree $C_{4,3}$. Therefore, the vertices in the last shell $g$ have degree $d_g=1$, and all vertices in other shells have degree $d_k=m$ $(k=0,1, \cdots, g-1)$ (the shell $k=0$ represents the single root vertex $O$). The number of nodes of the shell $k$ ($k=1, 2,\cdots, g$) is $N_{k}=m(m-1)^{k-1}$, while the total number of nodes and total number of edges in $C_{m,g}$ read
\begin{equation}\label{NodeC}
N=|V|=\frac{m(m-1)^g-2}{m-2}\,,
\end{equation}
\begin{equation}\label{EegeC}
|E|=N-1=\frac{m(m-1)^g-m }{m-2}\,,
\end{equation}
respectively.
The main difference between Bethe Lattice and Cayley tree is the fact that Bethe Lattice is infinite while Cayley tree finite, and $C_{m,g}$ can be considered as $BL_m$ confined in finite environments. All vertexes of $BL_m$ are equivalent, whereas vertexes of $C_{m,g}$ are divided into $g+1$ shells: the vertexes in different shells are not equivalent, while the vertexes in a same shell of $C_{m,g}$ are equivalent. In this paper, an arbitrary vertex in shell $k$ of $C_{m,g}$ is denoted by $v_k$ ($k=0,1,\dots,g)$.

\section{Survival probability 
} 
\label{sec: SP}
Let us consider target A and a moving trap B on Bethe Lattice or Cayley tree. The survival probability (SP) up to time $t$ for the target A is defined by
\begin{equation}
S(t)=1-\sum_{s=1}^t F(s),
\label{DefSP}
\end{equation}
where $F(t)$ is the first encounter probability (FEP), i.e., the probability that the two players arrive at the same site for the first time at time $t$. In order to find the effect of the target's movement on SP, we analyze FEP and SP, distinguishing between the cases with fixed and mobile target.


\subsection{Survival probability on Bethe Lattices}
\label{sec: Encounter_on_infinite_Cayley_trees}
First we consider SP for a target A in the presence of a moving trap B on Bethe Lattice $B_m$. Note that all vertexes of $B_m$ are equivalent. The SP for the target A is subject to the structure parameter $m$ of Bethe Lattice and the initial distance $L$ between the two walkers. 
Let $F_{Im}(t;m,L)$ and $S_{Im}(t;m,L)$ be the FEP and SP for immobile target, while $F_{M}(t;m,L)$ and $S_{M}(t;m,L)$ for mobile target, respectively.

Let the target A is immobile, and let $X_t$ be the distance between the two walkers at time $t$. Thus, we have $X_0=L$, and
\begin{equation}
X_{t+1}=\left\{\begin{array}{lll} X_t+1 & \text{with probability} & q=\frac{m-1}{m}, \\ X_t-1 & \text{with probability} &  p=\frac{1}{m},\end{array}\right.
\label{Xt_Im}
\end{equation}
for any $t\geq 0$, if  $X_t>0$. If $X_t=0$, the two players encounter at time $t$ and the game is over. Therefore, the first encounter processes for the two players on $B_m$ can be mapped into the first passage processes for biased random walk on semi-infinite line, which is modeled by $\{X_t, t\geq 0\}$, and $F(t;m,L)\equiv P\{X_t=0\}$. Here $P\{X_t=0\}$ is the probability that the walker, starting from $L$, first reaches $0$ at time $t$, on semi-infinite line. Recalling the result of $P\{X_t=0\}$ (e.g., see Ref.~\cite{Feller-1968}), we have $$\sum_tF_{Im}(t;m,L)=(\frac{1}{m-1})^L,$$ and for any integer $t$ $(t\geq L)$, 
\begin{equation}
F_{Im}(t;m,L)=\left\{ \begin{array}{ll}  0, & t<L,   \\\frac{L}{t}\tbinom{t}{(t+L)/2}p^{\frac{t+L}{2}}q^{\frac{t-L}{2}},  & t\geq L, \end{array} \right.
\label{PXt=0Im}
\end{equation}
where the binomial coefficient is zero if $t$ and $L$ are not of the same parity.

As for the case the  two walkers are mobile, let $Y_t$ be the distance between the two walkers at time $t$. Thus, we have $Y_0=L$, and
\begin{equation}
Y_{t+1}=\left\{\begin{array}{llc} Y_t+2 & w.p. & \frac{(m-1)^2}{m^2},\\ Y_t-2 & w.p. &  \frac{1}{m^2}, \\ Y_t & w.p. &  \frac{2m-2}{m^2},\end{array}\right.
\label{Yt_C2}
\end{equation}
for any $t>0$, if $Y_t>0$. If $Y_t=0$, the two players encounter at time $t$ and the game is over.
Therefore, the first encounter processes for the two walkers on $B_m$ can be modeled by $\{Y_t, t\geq 0\}$, and $F_{M}(t;m,L)=P\{Y_t=0\}$. 

For a random walk, modeled by $\{Y_t, t\geq 0\}$), at any step there is a nonzero probability $\frac{2m-2}{m^2}$ that the walker will stay in situ. Let $k$ be the total number of steps that the walker stay in situ, before it arrives at the site $0$ at time $t$. Then for the rest $t-k$ steps, the walker moves left or right and reach site $0$ at the $(t-k)$-th step. If we ignore the $k$ steps that the walker stay in situ, the process of other $t-k$ steps is similar to $\{X_t, t\geq 0\}$. The difference is in the probability $p=\frac{1}{(m-1)^2+1}$ (or $q=\frac{(m-1)^2}{(m-1)^2+1}$) that the walker moves to the left (or right), and here the walker jumps $2$ units at each step. The probability that the walker reaches $0$ at time $t-k$ can be exactly addressed by using a similar formula to Eq.~(\ref{PXt=0Im}).

If the walker for the first time arrives at site $0$ at time $t$, then $k$ steps that the walker stays in situ must happened in $t-1$ steps before time $t$, and $0\leq k\leq t-L/2$,  $t-k$ and $L/2$ are of the same parity\footnote{If $k>t-L/2$, we get $t-k<L/2$, the walker starting from site $L$ can not reach $0$ in $t-k$ steps and the probability that the walker reaches the site $0$ at time $t$ is $0$. Similarly, if $t-k$  and $L/2$ are not of the same parity, the walker starting from site $L$ can not reach $0$ in $t-k$ steps as well.}. Note that the probability that the walker stays in situ for $k$ times in $t-1$ steps before time $t$, reads
$$p_k=\tbinom{t-1}{k}(\frac{2m-2}{m^2})^k(\frac{m^2-2m+2}{m^2})^{n-k-1}.$$ By using the  whole probability formula, we get the FEP 
\begin{eqnarray}
& &F_{M}(t;m,L)=P\{Y_t=0, Y_n>0, 0<n<t\} \nonumber\\
&=&\sum_{k}\left\{p_k\frac{L/2}{t-k}\tbinom{t-k}{(t-k+L/2)/2}\left(\frac{1}{(m-1)^2+1}\right)^{\frac{t-k+L/2}{2}} \right.\nonumber\\
& &\left.\times\left(\frac{(m-1)^2}{(m-1)^2+1}\right)^{\frac{t-k-L/2}{2}}\right\},
\label{PYt=0}
\end{eqnarray}
for mobile target, where the sum runs over all possible $k\in\{k: 0\leq k\leq t-L/2$, $t-k$ and $L/2$ are of the same parity$\}$.

\begin{figure}
\begin{center}
{\mbox{\includegraphics[width=2in]{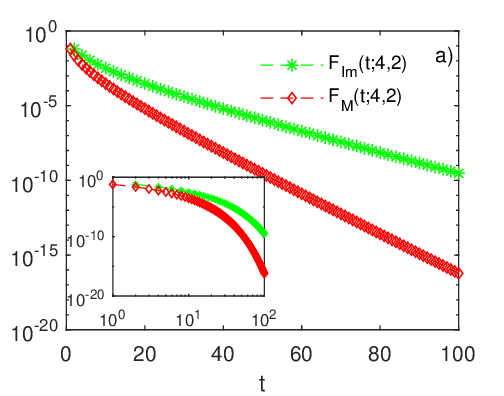}}}
{\mbox{\includegraphics[width=2in]{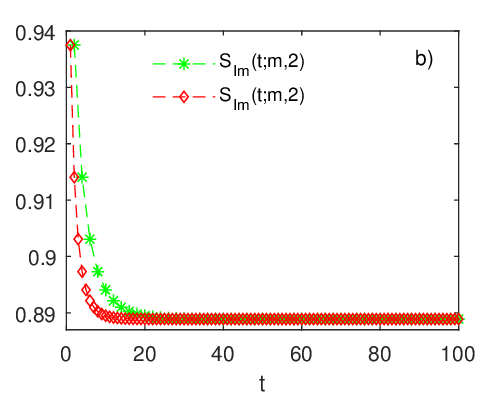}}}
\caption{(a) Plots of $F_{Im}(t;4,2)$ and $F_{M}(t;4,2)$ versus $t$. Data for  $F_{Im}(t;4,2)$ and $F_{M}(t;4,2)$ are obtained by evaluating the exact expression (\ref{PXt=0Im}) and(\ref{PYt=0}). Both $F_{Im}(t;4,2)$ and $F_{M}(t;4,2)$ decay exponentially with $t$ and the decay of $F_{M}(t;4,2)$ is faster than that of $F_{Im}(t;4,2)$. (b) Plots of  $S_{Im}(t;m,2)$ and $S_{M}(t;m,2)$ as  functions of $t$. Data for $S_{Im}(t;4,2)$ and $S_{M}(t;4,2)$ are obtained by inserting $F_{Im}(t;4,2)$ and $F_{M}(t;4, 2)$ into Eq.~(\ref{DefSP}). It shows $S_{Im}(t;m,L)\geq S_{M}(t;m,L)$ and $\lim \limits_{t\to \infty} {S_{Im}(t;m,L)}=\lim \limits_{t\to \infty}S_{M}(t;m,L)$.}
\label{fig:FEP_SP-BL}
\end{center}
\end{figure}

By comparing $F_{Im}(t;m,L)$ with $F_{M}(t;m,L)$, we find that both $F_{Im}(t;m,L)$ and $F_{M}(t;m,L)$ decay exponentially with $t$ and the decay of $F_{M}(t;m,L)$ is faster than $F_{Im}(t;m,L)$ (see FIG.~\ref{fig:FEP_SP-BL}~(a) for $m=4$ and $L=2$).

Inserting Eqs.~(\ref{PXt=0Im}) and (\ref{PYt=0}) into Eq.~(\ref{DefSP}), we obtain $S_{Im}(t;m,L)$ and $S_{M}(t;m,L)$. By comparing $S_{Im}(t;m,L)$ and $S_{M}(t;m,L)$, we find
\begin{equation}\label{COM_SP_BL}
S_{Im}(t;m,L)\geq S_{M}(t;m,L),
\end{equation}
and $$\lim \limits_{t\to \infty} {S_{Im}(t;m,L)}=\lim \limits_{t\to \infty}S_{M}(t;m,L)=1-(\frac{1}{m-1})^L,$$ see FIG.~\ref{fig:FEP_SP-BL}~(b) for $m=4$ and $L=2$. Just as expected, ``Pascal principle'' holds on Bethe Lattices, where the random walk is non-recurrent. 

\subsection{Survival probability   on  Cayley trees}
  \label{sec: FEP_on_finite_Cayley_trees}
Next we consider SP for the target A in the presence of a moving trap B on Cayley tree $C_{m,g}$ by means of numerical simulation. The FEP between the two walkers are also analyzed. Note that the initial positions of the two walkers have great effect on SP, and it is impossible to enumerate all the possible cases. In order to highlight the difference between SP on Bethe Lattice $B_m$ and SP on Cayley tree $C_{m,g}$, similarly to the problem analyzed in Sec.~\ref{sec: Encounter_on_infinite_Cayley_trees}, here we consider initial position of target A at the root of $C_{m,g}$ and initial distance $L$ between the two walkers. We also compare the SP of the target A for mobile target with the one for immobile target and disclose the effect of target's move on its SP.

\begin{figure}
\begin{center}
{\mbox{\includegraphics[width=2in]{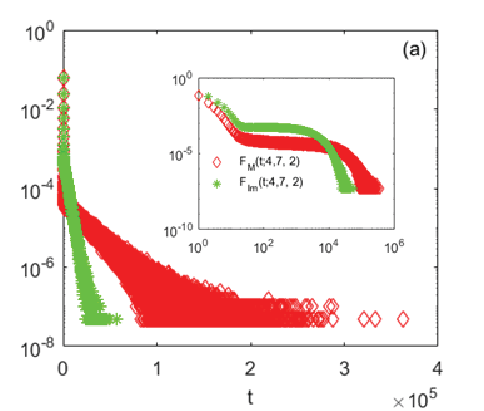}}}
{\mbox{\includegraphics[width=2in]{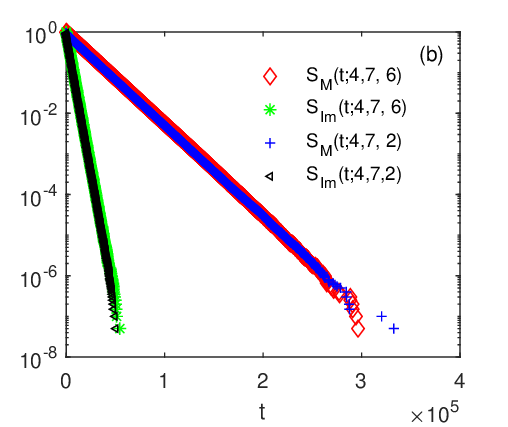}}}
\caption{(a) Plot of $F_{Im}(t;4,7, 2)$ and $F_{M}(t;4,7, 2)$ versus time $t$. The decay of $F_{Im}(t;4,7,2)$ and $F_{M}(t;4,7, 2)$ displays three regimes with different speed. The decays of $F_{Im}(t;4,7, 2)$ and $F_{M}(t;4,7, 2)$ are slower than those of $F_{Im}(t;4,2)$ and $F_{M}(t;4,2)$, as shown in FIG~\ref{fig:FEP_SP-BL} panel (a); (b) Plot of $S_{Im}(t;m,g,L)$ and $S_{M}(t;m,g,L)$ as function of $t$ for $m=4$, $g=7$, $L=2$ and $m=4$, $g=7$, $L=6$. For both cases, both $S_{Im}(t;m,g,L)$ and $S_{M}(t;m,g,L)$ decay exponentially with $t$, and the decay of $S_{Im}(t;m,g,L)$ is faster than that of $S_{M}(t;m,g, L)$, i.e., $S_{Im}(t;m,g, L)\leq S_{M}(t;m,g, L)$, which is not consistent with ``Pascal principle''. The data shown here have been obtained via numerical simulations and the sample, for each case, contains $10^7$ realizations.}
\label{fig:FEP_SP-CT}       %
\end{center}
\end{figure}

As in Sec.~\ref{sec: Encounter_on_infinite_Cayley_trees}, $F_{Im}(t;m,g,L)$ and $S_{Im}(t;m,g,L)$ denote the FEP and SP for immobile target, while $F_{M}(t;m,g,L)$ and $S_{M}(t;m,g,L)$ for mobile target, respectively. Contrary to the results on Bethe Lattice, we find that the decay of $F_{Im}(t;m,g,L)$ and $F_{M}(t;m,g, L)$ display three regimes with different speed, and the decays of $F_{Im}(t;m,g,L)$ and $F_{M}(t;m,g, L)$ are slower than the decay of $F_{Im}(t;m,L)$ and $F_{M}(t;m,L)$ (see FIG.~\ref{fig:FEP_SP-CT}~(a) for $m=4$, $g=7$, $L=2$). For initial position of the target A  at the root of Cayley tree $C_{m,g}$, we find
\begin{equation}\label{COM_SP_CT1}
S_{Im}(t;m,g,L)\leq S_{M}(t;m,g, L),
\end{equation}
which is opposite to the result on Bethe Lattice, as shown in Eq.~(\ref{COM_SP_BL}), and is not consistent with the ``Pascal principle''.  Therefore, in this sense, we cannot look on Bethe Lattices as the thermodynamic limit of Cayley trees. FIG.~\ref{fig:FEP_SP-CT}~(b) shows the plot of $S_{Im}(t;m,g,L)$ and $S_{M}(t;m,g,L)$ as functions of $t$ for $m=4$, $g=7$, $L=2$ and for $m=4$, $g=7$, $L=6$. Furthermore, Eq.~(\ref{COM_SP_CT1}) is robust and it also holds for initial position of the target A close to the root of the Cayley tree. 


\section{MFET and MFPT for two walkers on  Cayley trees}
  \label{sec: FET}

In what follows, the mean first encounter time (MFET) and some related quantities for two particles walking on Cayley trees, are evaluated and compared with corresponding quantities of mean first-passage time (MFPT) while one of the two particles is fixed. Different initial settings are considered. Here, the MFET is obtained by means of numerical simulation. 

\subsection{Definitions and method for numerical simulation}
\label{sec:Def_Methods}

For two walkers, A and B, performing simple random walk on Cayley tree $C_{m,g}$, the MFET between two walkers, $\textrm{MFET}_{(i_A, i_B)}$, is the mean time it takes the two walkers to be on the same site for the first time, after they leave their initial position $(i_A, i_B)$. For immobile target A, the MFET recovers the MFPT, $\textrm{MFPT}_{(i_A, i_B)}$, which is the mean time it takes the walker B starting from $i_B$ reaches $i_A$ for the first time.  

In order to get more synthetic information, one should analyze  the global-mean first-encounter time (GFET), defined by
\begin{equation}
\label{eq:GFET}
\textrm{GFET} := \sum_{i_A  \in V} \sum_{i_B  \in V} \pi_{i_A} \pi_{i_B} \textrm{MFET}_{(i_A, i_B)}.
\end{equation}
where $\pi_k = d_k / \sum_i d_i$, and $d_k$ is the degree of the arbitrary vertex $k$.

To find the effect of target's initial position $i_A$ on GFET, one can also consider GFET for fixed $i_A$, defined by
\begin{equation}
\label{eq:GFET_rA}
\textrm{GFET}_{i_A} := \sum_{i_B \in V}  \pi_{i_B}  \textrm{MFET}_{(i_A, i_B)}.
\end{equation}
For fixed target A, $\textrm{GFET}$ recovers the global-mean first-passage time (GFPT), defined by
\begin{equation}
\label{eq:GFPT}
\textrm{GFPT} := \sum_{i_A \in V} \sum_{i_B \in V} \pi_{i_A} \pi_{i_B} \textrm{MFPT}_{(i_A, i_B)},
\end{equation}
and 
\begin{equation}
\label{eq:GFPT_rA}
\textrm{GFPT}_{i_A} := \sum_{i_B \in V}  \pi_{i_B}  \textrm{MFPT}_{(i_A, i_B)}.
\end{equation}
\newline

In order to simulate a random walk on graph, the method for labeling all the vertexes and links of the graph is employed~\cite{Liu-2004-book, Amdjadi-2008-cnsns}, and the use of adjacency matrix is a common choice. However, it needs huge memory units when the size of the graph is large. As an alternative of adjacency matrix representation of the Cayley tree, here we introduce a new method, where the vertexes of Cayley tree $C_{m,g}$ are labeled by integers with base $m$, and the links between the vertexes can be found in the labels of the vertexes.

\emph{Labeling of the vertexes.} First, we label the root of the Cayley tree by integer $0$. Then, for any other vertex of Cayley tree $C_{m,g}$, it must be a vertex, denoted by $v_k$, in shell $k$ $(k=1,2,\dots,g)$, and there is an unique path $\{v_0, v_1, v_2,\dots, v_k\}$ from the root $v_0$ to the vertex $v_k$. Therefore, we can label the vertex by the unique path. The method is as follows. For any non-leaf vertex of Cayley tree $C_{m,g}$, we label its children sequentially with integers starting from $1$. Then the path $\{v_0, v_1, v_2,\dots, v_k\}$ (and vertex $v_k$) can be labeled by a sequence $\{i_1,i_2,\dots,i_k\}$, where $i_j$ represents the label of $v_j$ as a child of $v_{j-1}$. It is easy to find that the vertex labeled by $\{i_1,i_2,\dots,i_k\}$ is a child of vertex $\{i_1,i_2,\dots,i_{k-1}\}$ and there is a link between them. Note that the root of Cayley tree $C_{m,g}$ has $m$ children vertexes, and all other non-leaf vertexes have $m-1$ children vertexes. We have $1\leq i_1\leq m$ and $1\leq i_j\leq m-1$ for $j>1$. Therefore, the sequence $\{i_1,i_2,\dots,i_k\}$ can be represented by an integer with base $m$, which is equal to $\sum_{j=1}^k (i_j\times m^{k-j})$. Any vertex of Cayley tree $C_{m,g}$ can be labeled by an integer with base $m$. The label of its parents vertex can be obtained by removing the lowest bit of the integer, and its child vertex can be obtained by adding one more bit after the lowest bit of the integer.

\emph{Simulation of random walk.} For two walkers walking on $C_{m,g}$, we introduce two integers with base $m$, $p_A$ and $p_B$, to represent the current positions for the walkers. $p_A$ and $p_B$ change at every step. We remove the lowest bit of the integer when the walker moves to its parents vertex and we add one more bit after the lowest bit of the integer when the walker moves to one of its children vertexes. If $p_A=p_B$, encounter happens and the process ends. The total number of bits of both integers increases with the increase of the size $N$, and we should replace the two integers by two integer arrays when size $N$ is huge. However, we just need $O(log(N))$ memory units. By comparing with the adjacency matrix, which needs $O(N^2)$ memory units, the memory requirements of our method are negligible, and the difficulties for memory shortage are solved. We note that the method presented here can also be used on other regular hierarchical networks.

\subsection{Walkers starting from a same site}
\label{sec:MFET_the_same_site}

 Next, we consider a system of two walkers initially set at the same site ($i_A=i_B$), distinguishing between the cases with fixed and mobile target. Note that all nodes in the same shell of $C_{m,g}$ are equivalent and $C_{m,g}$ has just $g+1$ different shells. There are only $g+1$ different settings for the initial position of the two walkers. We analyze the MFET for $i_A=i_B=v_k$, $\textrm{MFET}_{{i_A=i_B=v_k}}$, where $v_k$ represents an arbitrary vertex in the shell $k$ $(k=0,1,2,...g)$, and compare $\textrm{MFET}_{{i_A=i_B=v_k}}$ with $\textrm{MFPT}_{{i_A=i_B=v_k}}$, where $\textrm{MFPT}_{{i_A=i_B=v_k}}$ is the MFPT for $i_A=i_B=v_k$.

\begin{figure}
\begin{center}
{\mbox{\includegraphics[width=2in]{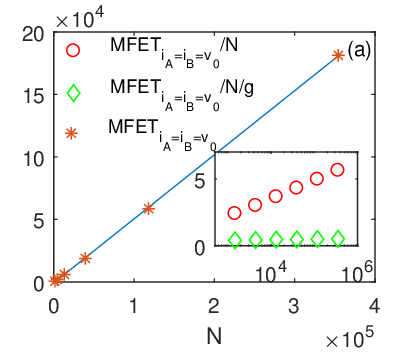}}}
{\mbox{\includegraphics[width=2in]{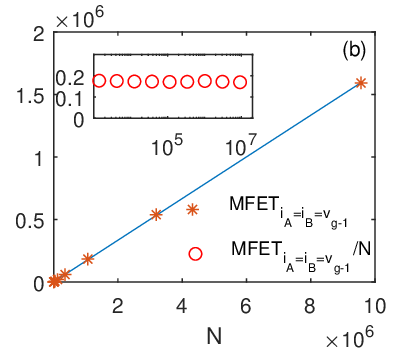}}}
\caption{(a) Main plot: plot of $\textrm{MFET}_{{i_A=i_B=v_0}}/g$ versus $N$ on Cayley trees with different size (i.e., $C_{4,g}$, $g=6,7,\dots,10$). $\textrm{MFET}_{{i_A=i_B=v_0}}/g$ scales linearly with $N$. Inset plot: $\textrm{MFET}_{{i_A=i_B=v_0}}/g/N$ and $\textrm{MFET}_{{i_A=i_B=v_0}}/N$ versus $N$. $\textrm{MFET}_{{i_A=i_B=v_0}}/N$ increases with $N$, $\textrm{MFET}_{{i_A=i_B=v_0}}/g/N$ is almost a constant as $N$ increases.
(b) Main plot: plot of $\textrm{MFET}_{{i_A=i_B=v_{g-1}}}$ versus  $N$ on Cayley trees with different size (i.e., $C_{4,g}$, $g=6,7,\dots,10$).  $\textrm{MFET}_{{i_A=i_B=v_{g-1}}}$ scales linearly with $N$. Inset plot:  $\textrm{MFET}_{{i_A=i_B=v_{g-1}}}/N$ versus $N$. $\textrm{MFET}_{{i_A=i_B=v_{g-1}}}/N$ is almost a constant as $N$ increases. The data shown here has been obtained via numerical simulations and, for every $g$, the sample contains $10^7$ realizations.}
\label{fig:3}       
\end{center}
\end{figure}

We analyze $\textrm{MFET}_{\substack{i_A=i_B=v_k}}$ $(k=0,1,2,...g)$ by means of numerical simulation, and disclose the relation between $\textrm{MFET}_{\substack{i_A=i_B=v_k}}$ and system size $N$. If $i_A=i_B=v_g$ (i.e., both walkers start from a leaf vertex of $C_{m,g}$), it follows that $\textrm{MFET}_{\substack{i_A=i_B=v_g}}=1$. If $i_A=i_B=v_{g-1}$, as shown in FIG.~\ref{fig:3}~(b), $\textrm{MFET}_{\substack{i_A=i_B=v_{g-1}}}/N$ is almost a constant as $N$ increases, and $\textrm{MFET}_{\substack{i_A=i_B=v_{g-1}}}$ scales linearly with the size $N$. Therefore, we get 
\begin{equation} \label{Mobilecasevg_1}
\textrm{MFET}_{\substack{i_A=i_B=v_{g-1}}}\sim N.
\end{equation}
If $i_A=i_B=v_{0}$ (i.e., both particles start from the root of $C_{m,g}$), as shown in FIG.~\ref{fig:3}~(a), $\textrm{MFET}_{\substack{i_A=i_B=v_0}}/N$ increases with $N$, $\textrm{MFET}_{\substack{i_A=i_B=v_0}}/g/N$ is almost a constant as $N$ increases, and $\textrm{MFET}_{\substack{i_A=i_B=v_0}}/g$ increases linearly with $N$. Therefore, we can state
\begin{equation} \label{Mobilecasev0}
\textrm{MFET}_{\substack{i_A=i_B=v_0}}\sim gN\sim Nlog(N).
\end{equation}

\begin{figure}
\begin{center}
{\mbox{\includegraphics[width=2in]{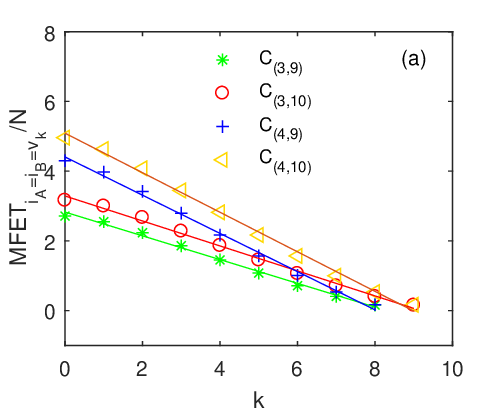}}}
{\mbox{\includegraphics[width=2in]{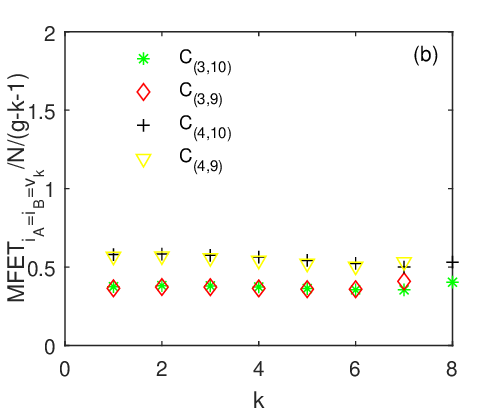}}}
\caption{(a) Plot of $\textrm{MFET}_{{i_A=i_B=v_k}}/N$ versus $k$ $(k=0, 1,2,\dots,g-1)$ on $C_{3,9}$, $C_{3,10}$, $C_{4,9}$ and $C_{4,10}$. $\textrm{MFET}_{{i_A=i_B=v_k}}/N$ decreases linearly as $k$ increases; (b) Plot of $(\textrm{MFET}_{{i_A=i_B=v_k}}/N)/(g-k-1)$ versus $k$ $(k=1,2,\dots,g-2)$ on $C_{3,9}$, $C_{3,10}$, $C_{4,9}$ and $C_{4,10}$. $(\textrm{MFET}_{{i_A=i_B=v_k}}/N)/(g-k-1)$ is almost a constant as $k$ increases. Therefore, $\textrm{MFET}_{{i_A=i_B=v_k}}\sim(g-k-1)N$. The data shown here have been obtained via numerical simulations and, for every $k$, the sample contains $10^7$ realizations.}
\label{fig:MFET_DN_g9_10}       
\end{center}
\end{figure}

In order to find the relation between $\textrm{MFET}_{{i_A=i_B=v_k}}$ and $N$ for all possible starting sites $v_k$ $(k=1,2,\dots,g-2)$, we also analyze the relation between $\textrm{MFET}_{\substack{i_A=i_B=v_k}}$ with $k$ on $C_{m,g}$. We find that $\textrm{MFET}_{\substack{i_A=i_B=v_k}}/N$ decreases linearly with the increase of $k$ (see FIG.~\ref{fig:MFET_DN_g9_10}~(a) for the results on $C_{3,9}$, $C_{3,10}$, $C_{4,9}$ and $C_{4,10}$), and $(\textrm{MFET}_{{i_A=i_B=v_k}}/N)/(g-k-1)$ is almost a constant as $k$ increases (see FIG.~\ref{fig:MFET_DN_g9_10}~(b) for the results on $C_{3,9}$, $C_{3,10}$, $C_{4,9}$ and $C_{4,10}$). Therefore,  $\textrm{MFET}_{{i_A=i_B=v_k}}\sim (g-k-1)N$ $(k=1,2,\dots,g-2)$. 
Recalling the scaling of $\textrm{MFET}_{\substack{i_A=i_B=v_{g-1}}}$ and $\textrm{MFET}_{\substack{i_A=i_B=v_0}}$, as shown in Eqs.~(\ref{Mobilecasevg_1}) and (\ref{Mobilecasev0}), we obtain
\begin{equation} \label{Mobilecase}
\textrm{MFET}_{\substack{i_A=i_B=v_k}}\sim (g-k)N,
\end{equation}
for $k=0,1,2,\dots,g-1$.

Let us now consider the MFPT in the case where one of the two particles is immobile and fixed at a given vertex $v_k$, while the other one is allowed to perform a random walk starting from the same site. In this case, the MFPT corresponds to the first return time of the mobile particle. Thus, one finds, see Ref.~\cite{Book-Lovasz-1993},
\begin{equation} \label{MFRT-ImMobilecase}
\textrm{MFPT}_{\substack{i_A=i_B=v_k}}=\frac{2|E|}{d_k}=
\left\{                 
  \begin{array}{ll}
   \frac{2[m(m-1)^g-m] }{(m-2)}, & k=g, \\
   \frac{2[m(m-1)^g-m ]}{m(m-2)}, & k<g.
  \end{array}
\right.                
 \end{equation}
Therefore $\textrm{MFPT}_{\substack{i_A=i_B=v_k}}\sim N$ for all $k$ $(0 \leq k \leq g )$.

\begin{figure}
\begin{center}
{\mbox{\includegraphics[width=2.in]{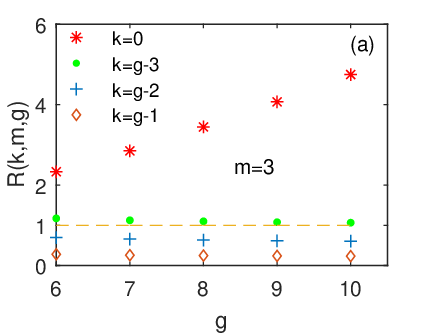}}}
{\mbox{\includegraphics[width=2.in]{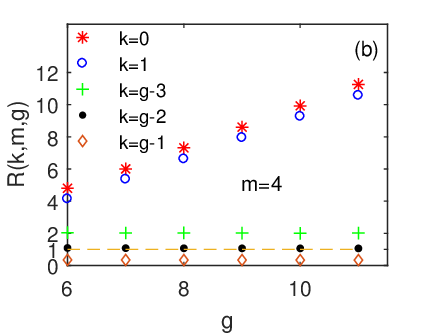}}}
{\mbox{\includegraphics[width=2in]{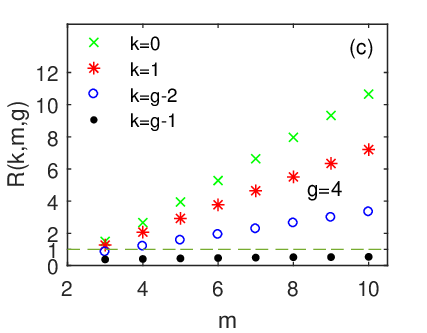}}}
{\mbox{\includegraphics[width=2in]{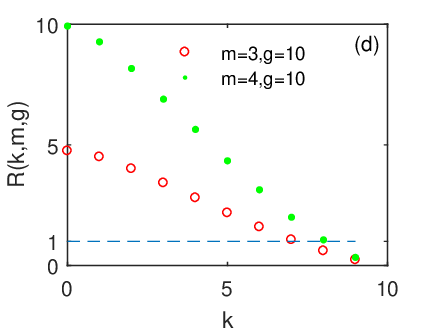}}}
\caption{(a) $R(k,m,g)$ versus $g$ for $m=3$ and different $k$ (i.e., $k=0, g-3, g-2, g-1$). For any $g$, $R(0,m,g)>R(g-3,m,g)>1$ and $R(g-1,m,g)<R(g-2,m,g)<1$, which presents evidence to support Eq.~(\ref{Com_MFRT-case1}); (b) $R(k,m,g)$ versus $g$ for $m=4$ for different $k$ (i.e., $k=0, 1, g-3, g-2, g-1$). For any $g$,  $R(0,m,g)>R(1,m,g)>R(g-3,m,g)>1$, $R(g-2,m,g)\approx 1$ and $R(g-1,m,g)<1$; (c) $R(k,m,g)$ as a function of $m$ ($m=3, 4, \dots, 10$) for $g=4$ and different $k$ (i.e., $k=0,1,g-2,g-1$). $R(k,m,g)$ increases with $m$, but $R(k,m,g)$ shows a little increase with $m$ while $k=g-1$. For $g=4$, $k=g-1$, $R(k,m,g)<1$ for all $m$; for $g=4$, $k=g-2$, $R(k,3,g)<1$, $R(k,3,g)>1$ for any $m>3$; for $g=4$, $k=1$ and $g=4$, $k=0$, $R(k,m,g)>1$ for any $m$; (d) $R(k,m,g)$ as a function of $k$ ($k=0,1,\dots,g-1$) for $m=3$, $g=10$ and $m=4$, $g=10$. $R(k,m,g)$ decreases as $k$ increases in both cases. In all four panels, $R(k,m,g)$ is obtained by dividing the numerical estimate for $\textrm{MFET}^{k}_{{i_A=i_B=v_k}}$ by the exact evaluation of $\textrm{MFPT}^{k}_{{i_A=i_B=v_k}}$ from (\ref{MFRT-ImMobilecase}). For every $(k,m,g)$, the sample of the simulation contains $10^7$ realizations.}
\label{fig:COM_MFET_MFRT}        
\end{center}
\end{figure}

In order to find the effect of the target's move on the encounter between the two walkers, one should analyze the ratio between  $\textrm{MFET}_{{i_A=i_B=v_k}}$ and $\textrm{MFPT}_{{i_A=i_B=v_k}}$. On Cayley tree $C_{m,g}$, let
\begin{equation}\label{Def_ratio}
R(k,m,g)\equiv \frac{ \textrm{MFET}_{{i_A=i_B=v_k}}}{\textrm{MFPT}_{{i_A=i_B=v_k}}},
\end{equation}
where $m\geq3$, $g\geq 1$ and $k=0,1,2,\dots,g$. Note that $R(k,m,g)<1$ for $k=g$. For $k<g$, 
we find that $R(k,m,g)$ monotonically increases with $m$ (see FIG.~\ref{fig:COM_MFET_MFRT}~(c), but $R(k,m,g)$ changes a little as $m$ increases while $k=g-1$, and $R(k,m,g)$ decreases monotonically as $k$ increases, see FIG.~\ref{fig:COM_MFET_MFRT}~(d).

For $m=3$, as shown in FIG.~\ref{fig:COM_MFET_MFRT}~(a), 
$R(0,m,g)>R(g-3,m,g)>1$ and $R(g-1,m,g)<R(g-2,m,g)<1$. Note that $R(k,m,g)$ decreases with $k$, and we can state
\begin{equation} \label{Com_MFRT-case1}
\left\{                 
  \begin{array}{ll}
   R(k,m,g)<1 & \textrm{if~~} k\geq g-2, \\
   R(k,m,g)>1 & \textrm{if~~} k\leq g-3.
  \end{array}
\right.                 
 \end{equation}
Similarly, for $m=4$ (see FIG.~\ref{fig:COM_MFET_MFRT}~(b), we get 
\begin{equation} \label{Com_MFRT-case2}
\left\{                 
  \begin{array}{ll}
   R(k,m,g)<1 & \textrm{if~~} k=g-1, \\
   R(k,m,g)\approx 1 & \textrm{if~~} k=g-2, \\
   R(k,m,g)>1 & \textrm{if~~} k\leq g-3.
  \end{array}
\right.                
 \end{equation}
Recalling $R(k,m,g)$ monotonically increases with $m$, for $m\geq 5$, we get
\begin{equation} \label{Com_MFRT-case3}
\left\{                 
  \begin{array}{ll}
   R(k,m,g)\approx R(k,4,g)<1 & \textrm{if~~} k=g-1,\\
   R(k,m,g)>R(k,4,g)>1 & \textrm{if~~} k\leq g-2.
  \end{array}
\right.                 
\end{equation}

All the results given by Eqs~(\ref{Com_MFRT-case1})-(\ref{Com_MFRT-case3}) are summarized in TABLE~\ref{tab}, which shows that $\textrm{MFET}_{i_A=i_B=v_k} > \textrm{MFPT}_{i_A=i_B=v_k}$ for $k\leq g-3$, i.e., the move of the target A slows down the encounter if the two walkers start from a site which is not close to the leaf-vertex of the Cayley tree. It is quite different with the result in Ref~\cite{Peng_Elena-PRE-2019}, which shows,
\begin{equation}
\label{Com:GFPT_gen}
\sum_{k \in V} \frac{d_k^2}{\sum d_i^2}  \textrm{MFPT}_{{i_A=i_B=v_k}}=2\sum_{k \in V} \frac{d_k^2}{\sum d_i^2}  \textrm{MFET}_{{i_A=i_B=v_k}},
\end{equation}
i.e., the move of the target A fastens the encounter if the same initial position of the two walkers are selected randomly. The reason is in the fact that there are so many leaf-vertexes in Cayley tree, where the move of target A fastens the encounter.

\begin{table}
\begin{center}
\begin{tabular}{{l|c|c}} \hline
$m$      &      $k$             & Encounter Times \\\hline
\multirow{2}*{$3$}
  &   $\geq g-2$  &  $\textrm{MFET}_{i_A=i_B=v_k} <  \textrm{MFPT}_{i_A=i_B=v_k}$   \\\cline{2-3}
 &   $\leq g-3$  &  $\textrm{MFET}_{i_A=i_B=v_k} > \textrm{MFPT}_{i_A=i_B=v_k}$   \\
\hline
\multirow{3}*{$4$}
   &   $\geq g-1$  &  $\textrm{MFET}_{i_A=i_B=v_k} <  \textrm{MFPT}_{i_A=i_B=v_k}$  \\\cline{2-3}
   &   $g-2$  &  $\textrm{MFET}_{i_A=i_B=v_k} \approx \textrm{MFPT}_{i_A=i_B=v_k}$  \\\cline{2-3}
   &   $\leq g-3$  &  $\textrm{MFET}_{i_A=i_B=v_k} > \textrm{MFPT}_{i_A=i_B=v_k}$   \\
\hline
\multirow{3}*{$\geq 5$}
   &   $\geq g-1$  &  $\textrm{MFET}_{i_A=i_B=v_k} <  \textrm{MFPT}_{i_A=i_B=v_k}$   \\\cline{2-3}
   &   $\leq g-2$  &  $\textrm{MFET}_{i_A=i_B=v_k} >  \textrm{MFPT}_{i_A=i_B=v_k}$  \\ \hline
\end{tabular}
\caption{The comparison between the MFET and MFPT on $C_{m,g}$ while $i_A=i_B=v_k$, with $m\geq 3, g\geq 3$, and $k=0, 1, \dots, g$.}\label{tab}
\end{center}
\end{table}


\subsection{Walkers starting from different but fixed sites}
\label{sec:MFET_different_sites}
We consider the case where A and B start from arbitrary fixed sites $i_A$ and $i_B$ on  $C_{m,g}$. We will analyze MFPT for immobile target A, and MFET for mobile target A. The MFPT is analyzed analytically based on the results obtained in our previous work~\cite{Peng-Stanley-2018-Jcp}, while the MFET is obtained numerically via simulations. Note that the initial position $(i_A, i_B)$ of the two walkers has great effect on MFPT and MFET. There are so many different choices of the pair $(i_A, i_B)$, and, thus, we can not enumerate all the possible choices. We just analyze the case with initial position of one of the walkers at the root.

We analyze the MFPT from $i_B$ to $i_A$, i.e., $\textrm{MFPT}_{(i_A, i_B)}$. By exploiting the connection between MFPT and effective resistance (see, e.g., Ref.~\cite{Peng_Elena-PRE-2019, Peng_2014-12-JSTAT,Peng-Stanley-2018-Jcp}), we get 
\begin{equation}
\textrm{MFPT}_{(i_A, i_B)} =|E|(L_{i_A,i_B}+W_{i_A}-W_{i_B}),
\label{MFPTj-i}
\end{equation}
where $L_{x,y}$ denotes the shortest-path length between vertex $x$ and $y$, $|E|=N-1$ is the total number of edges, and $W_y$ is defined as 
\begin{equation}
  W_y=\sum_{x \in V}{\frac{d_x}{2|E|}L_{x,y}}.
\label{Wy}
\end{equation}
Therefore, if $(L_{i_A,i_B}+W_{i_A}-W_{i_B})\sim const$, 
\begin{equation}
\textrm{MFPT}_{(i_A, i_B)}\sim N,
\label{MFPT_Case1}
\end{equation}
and if $(L_{i_A,i_B}+W_{i_A}-W_{i_B})\sim g\sim log(N)$,
\begin{equation}
\textrm{MFPT}_{(i_A, i_B)}\sim Nlog(N).
\label{MFPT_Case2}
\end{equation}

In particular, if $i_A=v_0$ and $i_B=v_k$, $(k\neq 0)$, 
one finds $L_{i_A,i_B}=k$, $W_{i_A}-W_{i_B}=-k+\frac{2(m-1)}{m(m-2)}-\frac{2(m-1)^{1-k}}{m(m-2)}$ (see Ref.~\cite{Peng-Stanley-2018-Jcp}). Therefore, $(L_{i_A,i_B}+W_{i_A}-W_{i_B})\sim const$, and for any $k=1,2,\dots,g$, 
\begin{equation}
\textrm{MFPT}_{(v_0, v_k)}\sim N.
\label{MFPT_Case3}
\end{equation}
Contrary to this, if $i_B=v_0$ and $i_A=v_k$, , $(k\neq 0)$, 
$(L_{i_A,i_B}+W_{i_A}-W_{i_B})\approx 2k$, and for any $k=1,2,\dots,g$,
\begin{equation}
\textrm{MFPT}_{(v_k, v_0)}\sim Nk.
\label{MFPT_Case4}
\end{equation}
Particularly,  $\textrm{MFPT}_{(v_g, v_0)}\sim Ng\sim Nlog(N)$.

\begin{figure}
\begin{center}
{\mbox{\includegraphics[width=2in]{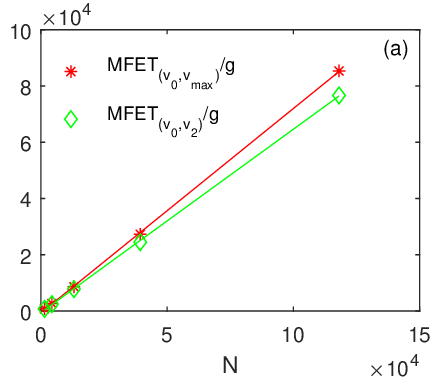}}}
{\mbox{\includegraphics[width=2in]{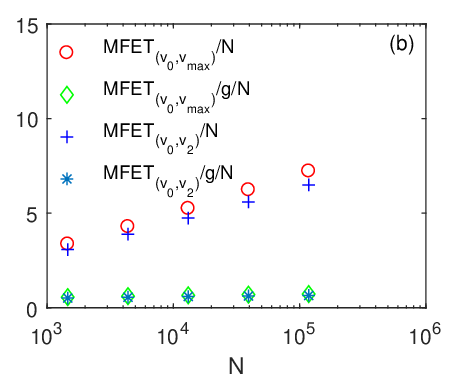}}}
\caption{(a) $\textrm{MFET}_{(v_{0},v_{2})}/g$ and $\textrm{MFET}_{(v_{0},v_{max})}/g$ as function of $N$ on Cayley trees with different size (i.e., $C_{m,g}$, $g=6,7,\dots,10$). Both $\textrm{MFET}_{(v_0, v_2)}/g$ and $\textrm{MFET}_{(v_0, v_{max})}/g$ increase linearly as $N$ increases; (b) Plot of $\textrm{MFET}_{(v_0, v_2)}/N$, $\textrm{MFET}_{(v_0, v_{max})}/N$, $\textrm{MFET}_{(v_0, v_2)}/g/N$ and $\textrm{MFET}_{(v_0, v_{max})}/g/N$ versus $N$ on Cayley trees with different size (i.e., $C_{m,g}$, $g=6,7,\dots,10$). Both $\textrm{MFET}_{(v_0, v_2)}/N$ and $\textrm{MFET}_{(v_0, v_{max})}/N$ increase with $N$, while both $\textrm{MFET}_{(v_0, v_2)}/g/N$ and $\textrm{MFET}_{(v_0, v_{max})}/g/N$ are almost constant as $N$ increases. The data shown here has been obtained via numerical simulations and, for every $k$, the sample contains $10^7$ realizations.}
\label{fig:MFET-scale} 
\end{center}
\end{figure}

Now we turn our analysis to $\textrm{MFET}_{(i_A, i_B)}$, which is the MFET in case the two walkers start from two different sites $i_A$, $i_B$. Here we just analyze the cases while the initial position for one of the two walkers is at the root. For any even $k$, we get
\begin{equation}
\textrm{MFET}_{(v_0, v_k)}=\textrm{MFET}_{(v_k, v_0)}\sim Nlog(N),
\label{MFET_Case}
\end{equation}
by means of numerical simulation. Part of the results are shown in FIG.~\ref{fig:MFET-scale}, which represent the numerical results for two walkers with a minimum distance (i.e., $i_A=v_0$, $i_B=v_2$), and maximum distance (i.e., $i_A=v_0$, $i_B=v_{max}$, with $max=g$ if $g$ is even and $max=g-1$ if $g$ is odd). As shown in FIG.~\ref{fig:MFET-scale}~(a), both $\textrm{MFET}_{(v_0, v_2)}/g$ and $\textrm{MFET}_{(v_0, v_{max})}/g$ increase linearly with the size $N$; and in panel (b), we find that both $\textrm{MFET}_{(v_0, v_2)}/N$ and $\textrm{MFET}_{(v_0, v_{max})}/N$ increase as $N$ increases, whereas both $\textrm{MFET}_{(v_0, v_2)}/g/N$ and $\textrm{MFET}_{(v_0, v_{max})}/g/N$ are almost constant as $N$ increases. Therefore, Eq,~(\ref{MFET_Case}) holds while $i_B=v_2$ and $i_B=v_{max}$. Note that MFET increases with the distance between the initial position of the two walkers. We can state that Eq.~(\ref{MFET_Case}) holds for any even $k$. For odd $k$, the two walkers can not encounter for ever because Cayley tree is a bipartite graph.

\begin{figure}
\begin{center}
{\mbox{\includegraphics[width=2in]{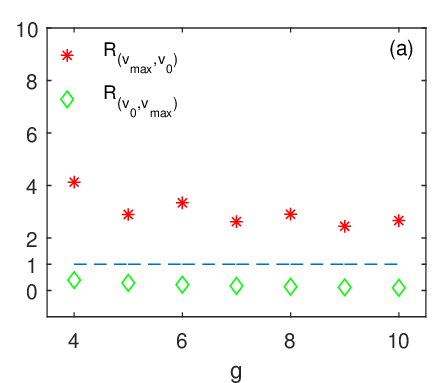}}}
{\mbox{\includegraphics[width=2in]{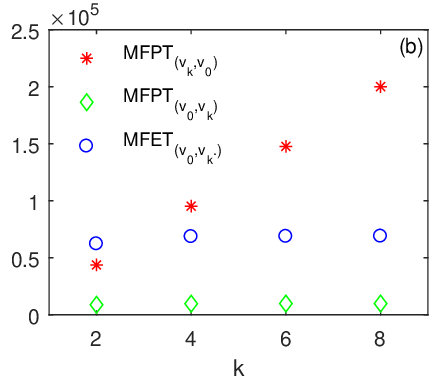}}}
\caption{(a) The ratio $R_{(v_{max},v_{0})}=\frac{\textrm{MFPT}_{(v_{max},v_{0})}}{\textrm{MFET}_{(v_{max},v_{0})}}$ and $R_{(v_{0},v_{max})}=\frac{\textrm{MFPT}_{(v_{0},v_{max})}}{\textrm{MFET}_{(v_0,v_{max})}}$ as function of $g$ on Cayley trees $C_{m,g}$ ($m=4$, $g=4,5,\dots,10$). Data for MFET are obtained via numerical simulations while for MFPT are obtained by evaluating the exact expression~(\ref{MFPTj-i}). Results show that $R_{(v_{max},v_{0})}>1$ and $R_{(v_{0},v_{max})}<1$ for all $g$, which confirm the correctness of Eqs.~(\ref{Com_MFET_MFPT_C1}) and (\ref{Com_MFET_MFPT_C2}); (b) $\textrm{MFPT}_{(v_0,v_{k})}$, $\textrm{MFPT}_{(v_k,v_{0})}$ and $\textrm{MFET}_{(v_0,v_{k})}$ versus $k$, and compared on Cayley tree $C_{4,8}$. Data for $\textrm{MFET}_{(v_0,v_{k})}$ are obtained via numerical simulations while for  $\textrm{MFPT}_{(v_0,v_{k})}$ and $\textrm{MFPT}_{(v_k,v_{0})}$ are obtained by evaluating the exact expression~(\ref{MFPTj-i}). Results show that $\textrm{MFPT}_{(v_0,v_{k})}<\textrm{MFET}_{(v_0,v_{k})}$ for all $k$,  whereas  $\textrm{MFPT}_{(v_k,v_{0})}>\textrm{MFET}_{(v_k,v_{0})}=\textrm{MFET}_{(v_0,v_{k})}$ for $k\geq 4$. They all confirm the correctness of Eqs.~(\ref{Com_MFET_MFPT_C1}) and (\ref{Com_MFET_MFPT_C2}).}
\label{fig:6}
\end{center}
\end{figure}

Comparing Eq.~(\ref{MFPT_Case3}) and Eq.~(\ref{MFET_Case}), as long as $N$ is large enough, we find
 \begin{equation}
\textrm{MFPT}_{(v_0, v_k)} < \textrm{MFET}_{(v_0, v_k)}.
 \label{Com_MFET_MFPT_C1}
 \end{equation}
It has been checked by numerical results as shown in FIG.~\ref{fig:6}. 
We find that Eq.~(\ref{Com_MFET_MFPT_C1}) is robust, and $\textrm{MFPT}_{(i_A, i_B)} < \textrm{MFET}_{(i_A, i_B)}$ while $i_A$ is close to the root of the Cayley trees (e.g. $i_A=v_0,v_1,v_2$). Therefore, the move of the target A slows down the encounter for a start site $i_A$ close to the root of the Cayley trees.

Our numerical results also show that, see FIG.~\ref{fig:6}
 \begin{equation}
\textrm{MFPT}_{(i_A, v_0)} > \textrm{MFET}_{(i_A, v_0)},
 \label{Com_MFET_MFPT_C2}
 \end{equation}
 while $i_A$ is close to the leaf-vertex of the Cayley trees (e.g. $i_A=v_{g-1},v_g$). Therefore, the move of the target A fastens the encounter for a start site close to the leaf-vertex of the Cayley trees. 


\subsection{One walker starting randomly}
\label{sec:A_F_B_R}
Here, we consider the case where the initial position of the target A is fixed at $i_A$, while the initial position of the trap B, $i_B$, is random and drawn from stationary distribution. We will analyze the GFET, which is obtained by averaging the MFET over all possible $i_B$, as defined in (\ref{eq:GFET_rA}), and then compare the GFET with GFPT, as defined in (\ref{eq:GFPT_rA}). The GFET is obtained numerically for different choices of $i_A$, while the GFPT is obtained by referring to our previous work~\cite{Peng-Stanley-2018-Jcp}.

Note that all nodes in the same shell of $C_{m,g}$ are equivalent. There are only $g+1$ different kinds of choices for the initial position of the target A. Let $v_k$ be an arbitrary vertex in the shell $k$ ($k=0,1,2,\dots,g$). We will analyze $\textrm{GFET}_{i_A=v_k}$ and $\textrm{GFPT}_{i_A=v_k}$, which are GFET and GFPT while $i_A$ is fixed at $v_k$, respectively.

\begin{figure}
\begin{center}
{\mbox{\includegraphics[width=2in]{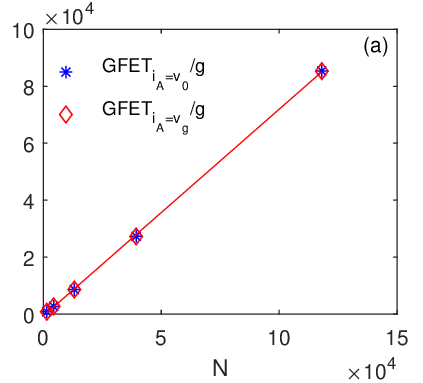}}}
{\mbox{\includegraphics[width=2in]{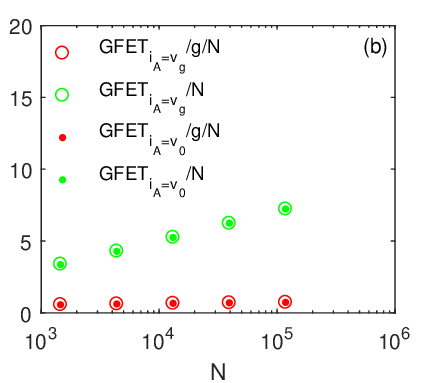}}}
\caption{(a) Plot of $\textrm{GFET}_{i_A=v_k}/g$ ($k=0,g$) versus $N$ on Cayley trees $C_{m,g}$ with $m=4$, $g=6,7,\dots,10$. Both $\textrm{GFET}_{i_A=v_0}/g$ and $\textrm{GFET}_{i_A=v_g}/g$ increase linearly with $N$, and $\textrm{GFET}_{i_A=v_0}/g \approx \textrm{GFET}_{i_A=v_g}/g$, which means that the selection of $i_A$ has a little effect on $\textrm{GFET}_{i_A=v_k}$; (b) Plot of $\textrm{GFET}_{i_A=v_k}/N$ and $\textrm{GFET}_{i_A=v_k}/g/N$ ($k=0,g$) versus $N$ on Cayley trees $C_{m,g}$ with $m=4$, $g=6,7,\dots,10$. Both $\textrm{GFET}_{i_A=v_0}/N$ and $\textrm{GFET}_{i_A=v_g}/N$  increase with $N$, whereas $\textrm{GFET}_{i_A=v_0}/g/N$ and $\textrm{GFET}_{i_A=v_g}/g/N$ are almost constant as $N$ increases. The data shown here have been obtained via numerical simulations and, for every initial position $i_A$, the sample contains $10^7$ realizations.}
\label{fig:GFETk=0_g}
\end{center}
\end{figure}

For $\textrm{GFPT}_{i_A=v_k}$, we find that $\textrm{GFET}_{i_A=v_k}/g$ increases linearly with the size $N$, and $\textrm{GFET}_{i_A=v_k}/N$ increases as $N$ increases, whereas $\textrm{GFET}_{i_A=v_k}/g/N$ is almost constant as $N$ increases (see FIG.~\ref{fig:GFETk=0_g} for $i_A=v_0$ and $i_A=v_g$ on Cayley trees $C_{m,g}$ with $m=4$, $g=6,7,\dots,10$). Therefore, we can state that 
\begin{equation}
\textrm{GFET}_{i_A=v_k}\sim gN\sim Nlog(N),
\label{GFET_Case1}
\end{equation}
for $k=0,1,2,\dots,g$.

 We also find, for any different $i,j$, $\textrm{GFET}_{i_A=v_i}\approx \textrm{GFET}_{i_A=v_j}$, which means that the initial position $i_A$ has a little effect on $\textrm{GFET}_{i_A=v_k}$ (see FIG.~\ref{fig:GFETk=0_g}, where the signs of $\textrm{GFET}_{i_A=v_0}$ are overlapped with those of $\textrm{GFET}_{i_A=v_g}$, with $v_0$ and $v_g$ being two sites with largest difference), and we can look on $\textrm{GFET}_{i_A=v_k}$ as a quantity which is independent of $k$.



As for $\textrm{GFPT}_{i_A=v_k}$, for an immobile target A,  Refs.~\cite{Wu-Lin-Zhang-2012, Peng-Stanley-2018-Jcp} shows that 
\begin{eqnarray}
\textrm{GFPT}_{i_A=v_k}&=&(m-1)^{g}[\frac{2km}{m-2}-\frac{2(m-1)}{(m-2)^2}]\nonumber \\
                 & &+\frac{4(m-1)^{g-k+1}}{(m-2)^2}-\frac{m(3m^2-8m+8)}{(m-2)^3}\, \nonumber \\
                  &\approx&2kN-\frac{2(m - 1)+4(m - 1)^{1-k}}{m(m - 2)}N, \label{MGFPTk}
\end{eqnarray}
for any $k=0,1,2,\dots,g$. 
Therefore, $\textrm{GFPT}_{{i_A=v_k}}$ increases monotonically in respect to $k$, and the initial position of the target A has a great effect on the GFPT. Particularly, $\textrm{GFPT}_{{i_A=v_k}}\sim N$ for $k=0$, and $\textrm{GFPT}_{{i_A=v_k}}\sim Nlog(N)$ for $k=g$. Hence, $r(k, m, g)$ increases with $k$, where
\begin{equation}
  r(k, m, g)\equiv\frac{\textrm{GFPT}_{i_A=v_k}}{\textrm{GFET}_{i_A=v_k}}
\end{equation}
with $m$ and $g$ being parameters of the Cayley tree $C_{m,g}$.

\begin{figure}
\begin{center}
{\mbox{\includegraphics[width=2in]{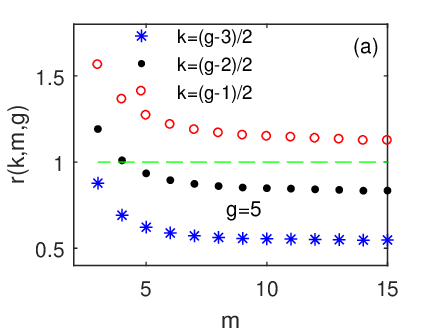}}}
{\mbox{\includegraphics[width=2in]{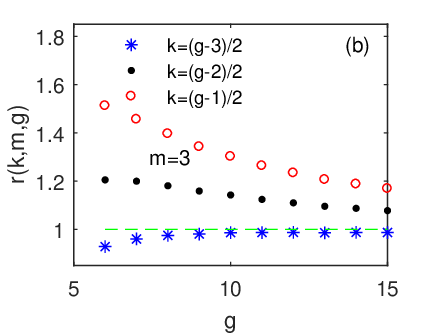}}}
{\mbox{\includegraphics[width=2in]{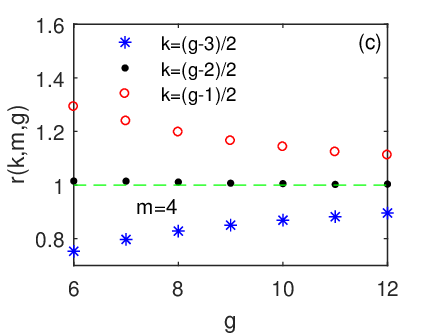}}}
{\mbox{\includegraphics[width=2in]{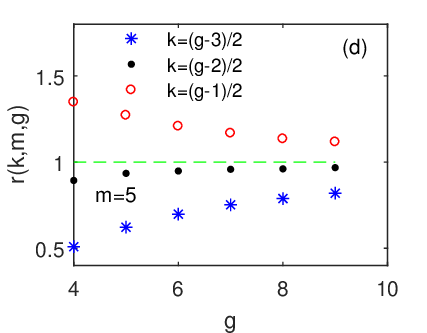}}}
\caption{(a) $r(k,m,g)$ versus $m$ for $g=5$ and different $k$ (i.e., $k=\frac{g-3}{2}, \frac{g-2}{2}, \frac{g-1}{2}$). $r(k, m, g)$ decreases as $m$ increases and expression~(\ref{Com_GMFET-GFPT}) holds. It also informs us that $r(\frac{g-2}{2},m,g)>1$ while $m=3$, $r(\frac{g-2}{2},m,g)\approx 1$ while $m=4$, and $r(\frac{g-2}{2},m,g)<1$ while $m\geq 5$, which give us an evidence to support of expression~(\ref{Com_GMFET-case2});
(b)-(d) $r(k,m,g)$ versus $g$ for $m=3$, $m=4$, $m=5$, and different choices of $k$, respectively. Expressions (\ref{Com_GMFET-GFPT}) and (\ref{Com_GMFET-case2}) hold for all considered cases. In all four panels, $r(k,m,g)$ is obtained by dividing the exact evaluation of $\textrm{GFPT}_{{i_A=v_{k}}}$ from Eq.~(\ref{MGFPTk}) by the numerical estimates for any ${\textrm{GFET}_{i_A=v_j}}$, $j=0,1,2,\dots,g$ and, for every triple $(k,m,g)$ the sample of the simulation contains $10^7$ realizations.}
\label{fig:Com_GFPTk_GFETk}        
\end{center}
\end{figure}

As shown in FIG.~\ref{fig:Com_GFPTk_GFETk}, we also find that $r(k, m, g)$ decreases as $m$ increases, see panel (a), and
\begin{equation} \label{Com_GMFET-GFPT}
r(\frac{g-1}{2}, m, g)\geq 1\geq r(\frac{g-3}{2}, m, g),
\end{equation}
for any $m$ and $g$, see panels (a)-(d)\footnote{Note that $\textrm{GFET}_{i_A=v_k}$ can be considered as a quantity which is independent of $k$ and $\textrm{GFPT}_{{i_A=v_k}}$ can be exactly evaluated from Eq.~(\ref{MGFPTk}) for any real $k$. If $k$ is not an integer, $r(k, m, g)$ can also be obtained by dividing the exact evaluation of $\textrm{GFPT}_{{i_A=v_{k}}}$ from Eq.~(\ref{MGFPTk}) by the numerical estimates for any ${\textrm{GFET}_{i_A=v_j}}$, $j=0,1,2,\dots,g$, and then be used as a reference value.}. Therefore,
\begin{equation} \label{Com_GMFET-case1}
\left\{                 
  \begin{array}{ll}
   r(k,m,g)\leq r(\frac{g-3}{2}, m, g)\leq 1 & \textrm{if~~} k\leq \frac{g-3}{2}, \\
   r(k,m,g)\geq r(\frac{g-1}{2}, m, g)\geq 1 & \textrm{if~~} k\geq \frac{g-1}{2}.
  \end{array}
\right.                 
 \end{equation}



Thus, for any non-negative integer $k$, if $k\geq \frac{g-1}{2}$, the move of target A fastens the encounter between the two walkers,
\begin{equation}
\textrm{GFPT}_{{i_A=v_k}} \geq \textrm{GFET}_{{i_A=v_k}};
 \label{Com_GFET_GFPT_C1}
\end{equation}
and if $k\leq \frac{g-3}{2}$, the move of the target A slows down the encounter between the two walkers,
\begin{equation}
\textrm{GFPT}_{{i_A=v_k}} \leq \textrm{GFET}_{{i_A=v_k}}.
 \label{Com_GFET_GFPT_C2}
\end{equation}

As for $k= \frac{g-2}{2}$, from FIG.~\ref{fig:Com_GFPTk_GFETk}, we can find
 \begin{equation} \label{Com_GMFET-case2}
r(\frac{g-2}{2},m,g)\left\{                 
  \begin{array}{ll}
  > 1 & \textrm{if~~} m=3, \\
   \approx 1 & \textrm{if~~} m=4, \\
   < 1 & \textrm{if~~} m\geq 5.
  \end{array}
\right.                 
 \end{equation}
Therefore, if $k= \frac{g-2}{2}$ is a non-negative integer,
 \begin{equation} \label{Com_GFET_GFPT_C3}
\left\{                 
  \begin{array}{ll}
  \textrm{GFPT}_{{i_A=v_k}} > \textrm{GFET}_{{i_A=v_k}} & \textrm{if~~} m=3, \\
   \textrm{GFPT}_{{i_A=v_k}} \approx \textrm{GFET}_{{i_A=v_k}} & \textrm{if~~} m=4, \\
   \textrm{GFPT}_{{i_A=v_k}} < \textrm{GFET}_{{i_A=v_k}} & \textrm{if~~} m\geq 5.
  \end{array}
\right.                 
 \end{equation}

\subsection{Walkers starting randomly}
   \label{sec: A_B_Start_Random}

For completeness, we analyze walkers starting from sites $i_A$ and $i_B$ drawn independently from the stationary distribution $\Pi$ on a Cayley tree $C_{m,g}$. We analyze the GFET for a mobile target, and compare it with the GFPT for an immobile target. 
Note that the initial position $i_A$ has little effect on $\textrm{GFET}_{i_A=v_k}$. We have $\textrm{GFET}\approx \textrm{GFET}_{i_A=v_k}$. Recalling Eq.~(\ref{Com_GMFET-GFPT}), we get 
 \begin{equation}
\textrm{GFPT}_{{i_A=v_{\frac{g-3}{2}}}} < \textrm{GFET}<\textrm{GFPT}_{{i_A=v_{\frac{g-1}{2}}}},
 \label{Com_GFET_GFPT_gen}
 \end{equation}
where 
\begin{eqnarray}
\textrm{GFPT}_{{i_A=v_{\frac{g-1}{2}}}}&\approx& gN-\frac{m^2 - 2+4(m - 1)^{\frac{3-g}{2}}}{m^2 - 2m}N,\nonumber \label{MGFETU}
\end{eqnarray}
and
\begin{eqnarray}
\textrm{GFPT}_{{i_A=v_{\frac{g-3}{2}}}}&\approx& gN-\frac{3m^2 -4m+4(m - 1)^{\frac{5-g}{2}}}{m^2 - 2m}N.\nonumber \label{MGFETL}
\end{eqnarray}
Therefore, 
\begin{equation}
\textrm{GFET} \sim Nlog(N).
\label{GFET_Case}
\end{equation}

%



\begin{figure}
\begin{center}
\includegraphics[scale=0.8]{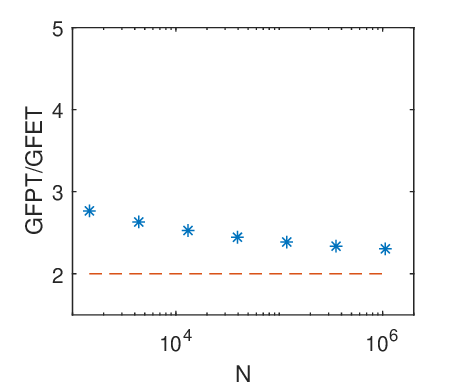}
\caption{The ratio $\textrm{GFPT}/ \textrm{GFET}$, obtained by dividing the exact evaluation of $\textrm{GFPT}$ from Eq.~(\ref{Kemeny_ER}) by the numerical estimates for $\textrm{GFET}$, versus $N$. $\textrm{GFPT}/ \textrm{GFET}>2$ and $\textrm{GFPT}/ \textrm{GFET}$ approaches $2$ as $N\rightarrow \infty$.}
\label{fig:Com_GFET_GFPT}       
\end{center}
\end{figure}

The GFPT can be rewritten as, see Ref.~\cite{Peng_Elena-PRE-2019}
 \begin{eqnarray} \label{GFPT_A_Im}
\textrm{GFPT}&=&|E|\sum_{y\in V}\frac{d_y}{2|E|}W_y,
\end{eqnarray}
where \cite{Peng-Stanley-2018-Jcp}
\begin{eqnarray}
\sum_{y\in V}\frac{d_y}{2|E|}W_y&=&\frac{2m(2gm^2-2m - 4gm  - m^2 + 2)(m - 1)^{2g}}{2(m - 2)^3 2|E|^2}\nonumber \\
  &+&\frac{  m(3m^2 + 4m - 4)(m - 1)^{g}-m^3}{2(m - 2)^3 2|E|^2}.
\label{WV}
\end{eqnarray}
Therefore,
\begin{eqnarray}\label{Kemeny_ER}
\textrm{GFPT}&=&\frac{2m(2gm^2 - m^2 -2m - 4gm  + 2)(m - 1)^{2g}}{2|E|(m - 2)^3 }\nonumber\\
    &+&\frac{m(3m^2 + 4m - 4)(m - 1)^{g}-m^3}{2|E|(m - 2)^3 }
    \nonumber\\
    &\approx& 2gN-\frac{m^2+2m-2}{m^2-2m}N\nonumber\\
    &\sim & Nlog(N).
 \end{eqnarray}
Comparing GFET and GFPT, as shown in Eqs.~(\ref{Com_GFET_GFPT_gen}) and (\ref{Kemeny_ER}), we get 
  \begin{equation}
\textrm{GFPT} > 2\textrm{GFET},
 \label{Com_GFET_GFPT}
 \end{equation}
and in the limit of large size, $N\rightarrow \infty$ (see FIG.~\ref{fig:Com_GFET_GFPT} for $m=4$),
\begin{equation}
\frac{\textrm{GFPT}}{\textrm{GFET}}\rightarrow 2.
 \end{equation}
Therefore the move of target A fastens the encounter between A and B.

\section{Conclusion} \label{sec:Conclusion}

Bethe Lattices are infinite regular trees with same degree of all vertexes, whereas Cayley trees are regular finite trees with same degree of all non-leaf vertexes. Bethe Lattices are often seen as the thermodynamic limit of Cayley trees. In this work we considered the first encounter problems between the target A (or a prey) and a moving trap B on Bethe Lattices and Cayley trees, distinguishing between the case with fixed and mobile target. The survival probability (SP) of the target A on both trees are evaluated. Results show, if we fix the distance between the two particles, whose initial positions are close to the root, the SP of a mobile target is less than or equal to the SP of an immobile target on Bethe Lattices, whereas the SP of a mobile target is greater than or equal to the SP of an immobile target on Cayley trees. Therefore, if the initial position of the pray is close to the root, in order to prolong survival time, on Bethe Lattices, it is better for the prey to stay still, whereas on Cayley trees -- the prey should move. In this sense, we can not consider the Bethe Lattices as thermodynamic limit of Cayley trees.

On Cayley trees, the MFET for a mobile target A is analyzed and compared with the MFPT for a fixed target A. Different initial position settings are considered and some interesting results are found. In general, if the initial position of the both particles are set randomly, the move of target A would, on average, speed up the encounter between them. However, if the initial position of target A is fixed at an arbitrary site in the shell $k$, and the initial position of the trap B is randomly drawn from the stationary distribution, two different results are observed. Namely, the move of the target A speeds up the encounter while $k\geq \frac{g-1}{2}$, whereas the move of the target A slows down the encounter while $k\leq \frac{g-3}{2}$. These findings would be helpful for prolonging the survival time of the target on Cayley trees. The target should move if its initial position is close to the root, but it should stay still if its initial position is close to the leaf-vertex. It is known, see Ref.~\cite{Peng_Elena-PRE-2019}, that the move of the target A would, on average, speed up the encounter, if the two walkers start from the same site on Cayley tree and if the same initial position of the two walkers are selected randomly. However, if we classify all the vertexes of Cayley tree $C_{m,g}$ into different shells according to its distance from the root, and let the two walkers both start from the same site in shell $k$, we find that the argument just holds for $k=g$ or $k=g-1$, and for all the other cases where $k=0,1,\dots,g-3$, the move of the target A slows down the encounter. If two friends, separating at a certain site in Cayley tree, want to meet again quickly, it is better one of the two friends to stay still if their initial position is not close to the leaf-vertex. All these findings would be helpful for optimizing the search process on regular branch structures in relation to the motion of the target and to the initial position setting of the target. The effect of the target's move on the encounter time on other structures is an interesting topic and worth for further analysis.

\section*{Acknowledgment}

The authors are grateful to Elena Agliari for valuable suggestions. JHP is supported by the National Natural Science Foundation of China (Grant No. 61873069 and 61772147) and the National Key R\&D Program of China (Grant No. 2018YFB0803604). TS was supported by the Alexander von Humboldt Foundation. \\

\appendix 



\begin{thebibliography}{10}
\expandafter\ifx\csname url\endcsname\relax
  \def\url#1{\texttt{#1}}\fi
\expandafter\ifx\csname urlprefix\endcsname\relax\def\urlprefix{URL }\fi
\expandafter\ifx\csname href\endcsname\relax
  \def\href#1#2{#2} \def\path#1{#1}\fi

\bibitem{Rupprecht-Benichou-2016-PRE}
J.-F. Rupprecht, O.~B\'{e}nichou, R.~Voituriez, Optimal search strategies of
  run-and-tumble walks, Phys. Rev. E 94 (2016) 012117.

\bibitem{Sims-Southall-Humphries-Hays-2009-Nature}
D.~W. Sims, E.~J. Southall, N.~E. Humphries, G.~C. Hays, {Scaling laws of
  marine predator search behaviour}, Nature 451 (2008) 1098--1102.

\bibitem{Volchenkov2011-cnsns}
D.~Volchenkov, Random walks and flights over connected graphs and complex
  networks, Commun. Nonlinear Sci. Numer. Simul. 16~(1) (2011) 21--55.

\bibitem{Viswanathan-Buldyrev-Havlin-Luz-Raposo-Stanley-1999-Nature}
G.~M. Viswanathan, S.~V. Buldyrev, S.~Havlin, M.~G.~E. da~Luz, E.~P. Raposo,
  H.~E. Stanley, {Optimizing the success of random searches}, Nature 401 (1999)
  911--914.

\bibitem{Magdziarz-Zorawik-2017-cnsns}
M.~Magdziarz, T.~Zorawik, Method of calculating densities for isotropic
  ballistic l\'evy walks, Commun. Nonlinear Sci. Numer. Simul. 48 (2017) 462 --
  473.

\bibitem{Luca2012-PRL}
A.~Luca, Collective predation and escape strategies, Phys. Rev. Lett. 109~(11)
  (2012) 118104.

\bibitem{Koza-Zbi-1998-PRE}
Z.~Koza, H.~Taitelbaum, Spatiotemporal properties of diffusive systems with a
  mobile imperfect trap, Phys. Rev. E 57~(1) (1998) 237--243.

\bibitem{Vot-Escuder-2018-PRE-Encounter}
F.~L. Vot, C.~Escudero, E.~Abad, S.~B. Yuste, Encounter-controlled coalescence
  and annihilation on a one-dimensional growing domain, Phys. Rev. E 98~(3).

\bibitem{Forrester-1999-JPA-Probability}
P.~J. Forrester, Probability of survival for vicious walkers near a cliff, J.
  Phys. A 22~(13) (1999) L609.

\bibitem{Yuste-Oshanin-2008-PRE}
S.~B. Yuste, G.~Oshanin, K.~Lindenberg, O.~Benichou, J.~Klafter, Survival
  probability of a particle in a sea of mobile traps: A tale of tails, Phys.
  Rev. E 78 (2008) 021105.

\bibitem{Oshanin-Vasilyev-2009-PNAS}
G.~Oshanin, O.~Vasilyev, P.~L. Krapivsky, J.~Klafter, Survival of an evasive
  prey, Proc. Natl. Acad. Sci. U.S.A. 106~(33) (2009) 13696--13701.

\bibitem{Szabo-1988-PRL-Diffusion}
A.~Szabo, R.~Zwanzig, N.~Agmon, Diffusion-controlled reactions with mobile
  traps, Phys. Rev. Lett. 61~(21) (1988) 2496--2499.

\bibitem{Schehr-2013-JSP-Reunion}
G.~Schehr, S.~N. Majumdar, A.~Comtet, P.~J. Forrester, Reunion probability of n
  vicious walkers: Typical and large fluctuations for large n, J. Stat. Phys.
  150~(3) (2013) 491--530.

\bibitem{Campari-Cassi-2012-PRE}
R.~Campari, D.~Cassi, Random collisions on branched networks: how simultaneous
  diffusion prevents encounters in inhomogeneous structures, Phys. Rev. E
  86~(1) (2012) 021110.

\bibitem{George-Patel-2018-XXX}
M.~George, R.~Patel, F.~Bullo, The meeting time of multiple random walks,
  https://arxiv.org/abs/1806.08843v1 (2018).

\bibitem{Moreau2003-PRE-Pascal}
M.~Moreau, G.~Oshanin, O.~B\'enichou, M.~Coppey, Pascal principle for
  diffusion-controlled trapping reactions, Phys. Rev. E 67~(2) (2003) 045104.

\bibitem{Chen-Sun-2012-XXX}
L.-C. Chen, R.~F. Sun, {The Pascal Principle for a Particle Among Sub-diffusive
  Mobile Traps}, http://arxiv.org/abs/1203.1389v1 (2012).

\bibitem{Tejedor-2011-JPA-Encounter}
V.~Tejedor, M.~Schad, O.~B\'enichou, R.~Voituriez, R.~Metzler, Encounter
  distribution of two random walkers on a finite one-dimensional interval, J.
  Phys. A 44~(39) (2011) 395005.

\bibitem{Holcman-Kupka-2009-JPA}
D.~Holcman, I.~Kupka, The probability of an encounter of two brownian particles
  before escape, J. Phys. A 42~(31) (2009) 1943--1953.

\bibitem{Chen-Sun-2014-JTP}
L.-C. Chen, R.~F. Sun, A monotonicity result for the range of a perturbed
  random walk, J. Theor. Probab. 27~(3) (2014) 997--1010.

\bibitem{Peng_Elena-PRE-2019}
J.~H. Peng, E.~Agliari, First encounters on combs, Phys. Rev. E 100 (2019)
  062310.

\bibitem{Chen-Chen-2011-EJP}
X.~Chen, D.~Chen, Some sufficient conditions for infinite collisions of simple
  random walks on a wedge comb, Electron. J. Probab. 16 (2011) 1341--1355.

\bibitem{Benichou-2015-PRL-Diffusion}
O.~B\'enichou, P.~Illien, G.~Oshanin, A.~Sarracino, R.~Voituriez, Diffusion and
  subdiffusion of interacting particles on comblike structures, Phys. Rev.
  Lett. 115~(22) (2015) 220601.

\bibitem{Chen-Bei-2008-SPL}
D.~Chen, W.~Bei, F.~Zhang, A note on the finite collision property of random
  walks, Statist. Prob. Lett. 78~(13) (2008) 1742--1747.

\bibitem{Elena-2014-PRE}
E.~Agliari, A.~Blumen, D.~Cassi, Slow encounters of particle pairs in branched
  structures, Phys. Rev. E 89~(5) (2014) 052147.

\bibitem{Agliari-2016-PRE-Two}
E.~Agliari, D.~Cassi, L.~Cattivelli, F.~Sartori, Two-particle problem in
  comblike structures, Phys. Rev. E 93~(5) (2016) 052111.

\bibitem{Ostilli2012-Physica-A}
M.~Ostilli, Cayley trees and bethe lattices: A concise analysis for
  mathematicians and physicists, Physica A 391~(12) (2012) 3417 -- 3423.

\bibitem{Cassi_1989-EPL}
D.~Cassi, Random walks on bethe lattices, Europhys. Lett. 9~(7) (1989)
  627--631.

\bibitem{Wu-Lin-Zhang-2012}
B.~Wu, Y.~Lin, Z.~Z. Zhang, G.~R. Chen, Trapping in dendrimers and regular
  hyperbranched polymers, J. Chem Phys. 137~(4) (2012) 044903.

\bibitem{Lin-Zhang-JCP-2013}
Y.~Lin, Z.~Z. Zhang, Influence of trap location on the efficiency of trapping
  in dendrimers and regular hyperbranched polymers, J. Chem Phys. 138~(9)
  (2013) 094905.

\bibitem{Feller-1968}
W.~Feller, An Introduction to Probability Theory and its Applications Vol I
  (3rd edition), John Wiley Sons, Inc, New York-London-Sydney, 1968.

\bibitem{Liu-2004-book}
J.~S. Liu, Monte Carlo Strategies in Scientific Computing, Springer-Verlag, New
  York, 2004.

\bibitem{Amdjadi-2008-cnsns}
F.~Amdjadi, Numerical simulation of reaction¨cdiffusion equations on spherical
  domains, Commun. Nonlinear Sci. Numer. Simul. 13~(8) (2008) 1592 -- 1595.

\bibitem{Book-Lovasz-1993}
L.~Lov{\'a}sz, Combinatorics: Paul Erdos Is Eighty, Keszthely, Hungary, 1993.

\bibitem{Peng-Stanley-2018-Jcp}
J.~H. Peng, G.~A. Xu, R.~X. Shao, C.~Lin, H.~E. Stanley, Analysis of
  fluctuations in the first return times of random walks on regular branched
  networks, J. Chem Phys. 149~(2) (2018) 024903.

\bibitem{Peng_2014-12-JSTAT}
J.~H. Peng, Mean trapping time for an arbitrary node on regular hyperbranched
  polymers, J. Stat. Mech. 2014~(12) (2014) P12018.

\end{thebibliography}

\end{document}